\documentstyle[12pt,epsfig]{article}
\newlength{\dinwidth}
\newlength{\dinmargin}
\newcommand{\ba}{\begin{array}}
\newcommand{\ea}{\end{array}}
\newcommand{\beq}{\begin{equation}}
\newcommand{\eeq}{\end{equation}}
\newcommand{\bea}{\begin{eqnarray}}
\newcommand{\eea}{\end{eqnarray}}

\def\bce{\begin{center}}
\def\ece{\end{center}}

\def\nonu{\nonumber}

\def\pa{\partial}
\def\al{\alpha}

\def\de{\delta}

\def\ep{\epsilon}
\def\vep{\varepsilon}

\def\th{\theta}

\def\la{\lambda}
\def\La{\Lambda}

 \def\Si{\Sigma}

\def\S{{\bf S}}

\newcommand{\th}{\widetilde{h}}
\newcommand{\tW}{\widetilde{W}}
\newcommand{\tD}{\widetilde{D}}
\newcommand{\dfrac}[2]{\displaystyle\frac{\mathstrut #1}{\mathstrut #2}}
\newcommand{\dsqrt}[1]{\sqrt{\mathstrut #1}}

\def\Kop{{\displaystyle \mathop{K}^{\circ}}{}}
\def\Gop{{\displaystyle \mathop{g}^{\circ}}{}}

\def\eps6{{\displaystyle \mathop{\epsilon}^{6}}{}}

\def\nab6{{\displaystyle \mathop{\nabla}^{6}}{}}

\begin{document}
\thispagestyle{empty}
\addtocounter{page}{-1}
\begin{flushright}
{\tt hep-th/0112010}\\
revised, Feb., 2002\\
\end{flushright}
\vspace*{1.3cm}
\centerline{\Large \bf An ${\cal N}=1$ Supersymmetric $G_2$-invariant Flow}
\vskip0.3cm
\centerline{\Large \bf in $M$-theory}
\vspace*{1.5cm} 
\centerline{{\bf Changhyun Ahn}$^1$ and {\bf Taichi Itoh}$^2$}
\vspace*{1.0cm}
\centerline{\it $^1$Department of Physics, 
Kyungpook National University, Taegu 702-701, Korea}
\vspace*{0.2cm}
\centerline{\it $^2$Department of Physics, 
Hanyang University, Seoul 133-791, Korea}
\vskip2cm
\centerline{\bf Abstract}
\vspace*{0.5cm}

It was found that deformation of $\S^7$ gives rise to renormalization 
group(RG)
flow from ${\cal N}=8$, $SO(8)$-invariant UV fixed point to ${\cal N}=1$,
$G_2$-invariant IR fixed point in four-dimensional gauged ${\cal N}=8$ 
supergravity. Also BPS supersymmetric 
domain wall configuration interpolated between these
two critical points. In this paper, we use the $G_2$-invariant 
RG flow equations for both scalar fields and domain wall amplitude  
and apply them to the nonlinear metric ansatz developed by de Wit, Nicolai 
and Warner some time ago. 
We carry out the $M$-theory lift of the $G_2$-invariant RG flow
through a combinatoric use of the four-dimensional RG flow equations
and eleven-dimensional Einstein-Maxwell equations.
The non-trivial $r$(that is the coordinate transverse to the 
domain wall)-dependence of vacuum expectation values makes 
the Einstein-Maxwell equations consistent not only 
at the critical points but also along the supersymmetric RG flow 
connecting two critical points. By applying an ansatz for an 
eleven-dimensional three-form gauge field with varying scalars, we discover 
an exact solution to the eleven-dimensional Einstein-Maxwell equations 
corresponding to the $M$-theory lift of the $G_2$-invariant RG flow.

\vspace*{\fill}

\noindent
------------------\\
E-mail addresses: ahn@knu.ac.kr (C. Ahn), 
taichi@hepth.hanyang.ac.kr (T. Itoh)

\baselineskip=18pt
\newpage
\renewcommand{\theequation}{\arabic{section}\mbox{.}\arabic{equation}}

\section{Introduction}
\setcounter{equation}{0}

By generalizing compactification vacuum ansatz
to the nonlinear level, solutions of the 11-dimensional
supergravity were obtained directly from constant scalar and pseudo-scalar
expectation values at the various {\it critical points} of the 
${\cal N}=8$ supergravity potential \cite{dnw}.
They were able to reproduce all known Kaluza-Klein solutions of 
11-dimensional supergravity except for 
$SU(3)\times U(1)$-vacuum\footnote{
The 11-dimensional embedding of $SU(3)\times U(1)$-invariant vacuum 
solution was recently found in \cite{cpw} as a stretched 7-ellipsoid.}: 
round $\S^7$, $SO(7)^{-}$-invariant parallelized $\S^7$, 
$SO(7)^{+}$-invariant 7-ellipsoid, $SU(4)^{-}$-invariant stretched $\S^7$, 
and $G_2$-invariant 7-ellipsoid. Among them the only round $\S^7$
and $G_2$-invariant 7-ellipsoid are stable and supersymmetric. 
One of the important features of de Wit-Nicolai theory \cite{dn} 
is that four-dimensional spacetime when dimensionally reduced from
11-dimensional supergravity
is warped \cite{dnw,dn87} 
by warp factor $\Delta(y)$. To be explicit, we have
11-dimensional metric 
\bea
ds_{11}^2 = ds_4^2 + ds_7^2 =
\Delta^{-1}(y) \,g_{\mu \nu}(x) \,d x^{\mu} d x^{\nu}
+ g_{mn}(y) \,dy^m dy^n.
\label{metric}
\eea 
Novelty of vacua with a non-trivial warp factor
is that they correspond to inhomogeneous deformations of $\S^7$.
For $SO(7)^{-}$ and $SU(4)^{-}$ solutions, the function $\Delta(y)$
is a constant while for $SO(7)^{+}$ and $G_2$ solutions 
it is a non-trivial function of internal coordinate $y$. 
The existence of warp factor is crucial for understanding of
the different relative scales of the 11-dimensional solutions
corresponding to the critical points in ${\cal N}=8$ gauged supergravity. 
The main result of \cite{dnw} was that it is possible to
write down the metric directly from 
the vacuum expectation values of the 
scalar and pseudo-scalar fields in 4-dimensional theory.
Therefore both internal metric and warp factor depend on
4-dimensional coordinate $x$ as well as $y$. 
With explicit dependence on $x$ and $y$, 
the inverse metric on $\S^7$, $g^{mn}(x, y)$, is given by 
\cite{dnw,dn87} 
\beq
g^{mn}(x, y) = 
\Delta (x, y)\Kop^{mIJ}(y) \Kop^{nKL}(y)\! 
\left[ u_{ij}^{\;\;\;IJ}(x) + v_{ijIJ}(x) \right]\!
\left[ {\overline{u}}^{\,ij}_{\;\;\;KL}(x) + 
{\overline{v}}^{\,ijKL}(x) \right] 
\label{internalmetric}
\eeq 
where $\Kop^{mIJ}(y)$ is the usual Killing vector on the unit round 
$\S^7$(See also (\ref{killing})) and warp factor $\Delta(x, y)$ is defined by
\bea
\Delta(x, y) = \dsqrt{\frac{\mbox{det}\, g_{mn}(x, y)}
{\mbox{det}\, \Gop_{mn}(y)}}
\label{warp}
\eea
where the metric $\Gop_{mn}(y)$ 
is that of the round $\S^7$. 
The 28-beins 
$u_{ij}^{\;\;\;IJ}(x)$ and $v_{ijIJ}(x)$ fields are $28 \times 28$
matrices and are given in terms of scalar and pseudo-scalar fields
in four-dimensional gauged supergravity. We denote 
the complex conjugations of
$u_{ij}^{\;\;\;IJ}(x)$ and $v_{ijIJ}(x)$ by
${\overline{u}}^{\,ij}_{\;\;\;IJ}(x)$ and $
{\overline{v}}^{\,ijIJ}(x)$ respectively.
Moreover, 
$SU(8)$ index pairs $[ij]$ and $SO(8)$ index pairs $[IJ]$ are
antisymmetrized.

By analyzing scalar potential in the de Wit-Nicolai theory, 
it was found in \cite{ar2} 
that the deformation of $\S^7$ gives rise to renormalization group(RG)
flow from ${\cal N}=8$, $SO(8)$-invariant ultraviolet(UV) fixed point
to ${\cal N}=1$, $G_2$-invariant infrared(IR) fixed point, via AdS/CFT
correspondence \cite{malda,witten,gubser}. 
The flow {\it interpolates} between both fixed points.
The critical points give four-dimensional $AdS_4$ vacua and 
preserve $G_2$ gauge
symmetry in the supergravity side. Having established the holographic
duals of both supergravity critical points and examined small 
perturbations around the corresponding fixed point field theories,
one can proceed the supergravity description of the RG flow
between the two fixed points. The supergravity scalars tell us
that what relevant operators in the dual field theory would
drive a flow to the fixed point in the IR. To construct 
the superkink corresponding to the supergravity description
of the nonconformal RG flow connecting two critical points
in $d=3$ conformal field theories, the form of 
a three-dimensional Poincare invariant metric but breaking
full conformal group invariance takes the form
\bea
g_{\mu \nu}(x) \,d x^{\mu} d x^{\nu} = e^{2A(r)} \,\eta_{\mu' \nu'}\,
d x^{\mu'} d x^{\nu'} + dr^2, \qquad \eta_{\mu' \nu'} = (-, +, +)
\label{domain}
\eea
characteristic of spacetime with a domain wall where $r$ is the
coordinate transverse to the wall(interpreted as an energy scale)
and $A(r)$ is the scale factor
in the four-dimensional metric.
By minimization of energy-functional, in order to get BPS domain-wall
solutions, one has to reorganize it into complete squares. By recognizing
that de Wit-Nicolai theory has particular property, the scalar
potential is written as the difference of two positive square terms,
one can construct energy-functional in terms of complete squares. 
In \cite{aw}, we have found that the first order differential
equations with respect to the $r$-direction for the varying scalar fields 
are the gradient flow equations of a superpotential 
defined on a restricted slice of complete scalar manifold.

In this paper, we find $M$-theory solutions that are holographic
duals of flows of the maximally supersymmetric ${\cal N}=8$
theory in three-dimensions. 
By using the RG flow equations for scalar fields and $A(r)$, 
implying that one can find the derivatives of these fields
with respect to $r$ explicitly, we generalize the 
scheme of \cite{dnw} and study several aspects of the embedding
of gauged ${\cal N}=8$ supergravity into 11-dimensional supergravity.
We will begin our analysis in Section 2 by 
summarizing relevant aspects of \cite{aw} in four-dimensional
gauged supergravity. In Section 3, we will investigate 
the lift of the four-dimensional solution to $M$-theory by solving 
11-dimensional Einstein-Maxwell equations. We will start with the metric 
(\ref{metric}) together with (\ref{internalmetric}), (\ref{warp}) 
and (\ref{domain}). Then we will apply 
an ansatz for 11-dimensional 3-form gauge field which is a natural extension
of the Freund-Rubin parametrization \cite{fr} but will be more complicated 
since we are dealing with non-constant vacuum expectation values. 
Recently,  
$M$-theory lift of the $SU(3) \times U(1)$-invariant RG flow was found 
in \cite{cpw,jlp} by applying the RG flow equations given in \cite{ap}.
Similar analysis in $AdS_5$ supergravity and its lift to
type IIB string theory were given in \cite{kw}. 
In Section 4, we will summarize our results and will discuss about future 
direction.

Throughout this paper, we will be using the metric convention 
$(-, +, \cdots, +)$. Our notation is that the $d=11$ coordinates
with indices $M, N, \cdots$ are decomposed into $d=4$ spacetime 
coordinates $x$ with indices $\mu, \nu, \cdots$ and
$d=7$ internal space coordinates $y$ with indices $m, n, \cdots$.
Denoting the 11-dimensional metric as $g_{MN}$ and the antisymmetric 
tensor fields as $F_{MNPQ} = 4\,\pa_{[M} A_{NPQ]}$, the bosonic
field equations are \cite{cjs}
\bea
R_{M}^{\;\;\;N} & = & \frac{1}{3} \,F_{MPQR} F^{NPQR}
-\frac{1}{36} \de^{N}_{M} \,F_{PQRS} F^{PQRS},
\nonu \\
\nabla_M F^{MNPQ} & = & -\frac{1}{(4!)^2} \,E \,\ep^{NPQRSTUVWXY}
F_{RSTU} F_{VWXY},
\label{fieldequations}
\eea
where the covariant derivative on $F^{MNPQ}$ in the second relation
is given by $E^{-1} \pa_M ( E F^{MNPQ} )$ together with elfbein determinant 
$E = \sqrt{-g_{11}}$. The 11-dimensional epsilon tensors with lower indices 
$\ep_{NPQRSTUVWXY}$ are purely numerical.

\section{The $G_2$-invariant holographic RG flow in 4 dimensions}
\setcounter{equation}{0}

The ungauged ${\cal N}=8$ supergravity \cite{cremmer} has  a local 
compact symmetry
of the action $H=SU(8)$ and a global non-compact symmetry of the
equations of motion $G=E_{7(+7)}$, of which the subgroup $L=SL(8, {\bf R})$
is a global symmetry of the action. 
An arbitrary element of the 133-dimensional Lie algebra of $E_{7(+7)}$
can be represented by a $56 \times 56$ matrix(four $28 \times 28$
block matrices)
\bea
\left( \begin{array}{ccc} 
\La_{IJ}^{\;\;\;KL} & \Si_{IJPQ} \\
 \overline{\Si}^{MNKL} & \overline{\La}^{MN}_{\;\;\;\;\;PQ}  
\end{array} \right)
\nonu
\eea
where the indices $I,J = 1, \cdots, 8$ are antisymmetric in pairs. 
The $H=SU(8)$ maximally compact subgroup of $E_{7(+7)}$ is generated by
the 63-dimensional diagonal subalgebra 
\bea
D(\La_I^{\;\;J}) = \left( \begin{array}{ccc} 
\underline{\La}_{IJ}^{\;\;\;KL} & 0 \\
 0 & \underline{\overline{\La}}^{MN}_{\;\;\;\;\;PQ}  
\end{array} \right), \qquad \underline\La=
\underline{\La}_{IJ}^{\;\;\;PQ}= \delta_{[I}^{\;\;\;[P} \La_{J]}^{
\;\;\;Q]}
\nonu
\eea
where
$\La_{I}^{\;\;J}$ is an $8 \times 8$, antihermitian trace-free
generator of $SU(8)$: $\La_{I}^{\;\;J}= - \overline{\La}^J_{\;\;I}, 
\La_{I}^{\;\;I}=0$. The 70 non-compact generators are parametrized
by the complex, self-dual antisymmetric tensors $\Si_{MNPQ}$  
that satisfy
\bea
 \overline{\Si}^{MNPQ} = (\Si_{MNPQ} )^{\ast} =
\frac{1}{24} \eta \ep^{IJKLMNPQ} \Si_{IJKL}
\nonu
\eea
where $\eta = \pm 1$ is an arbitrary phase, chosen as $+1$.
Then, $L=SL(8, {\bf R})$ is the real subgroup of $E_{7(+7)}$  given by
restricting the above 133 generators to the 28 generators of 
$SO(8) \subset SU(8)$, $\La_I^{\;\;J}(= \overline{\La}^I_{\;\;J})$ plus
the 35 real, self-dual antisymmetric tensors, $\Si_{IJKL}(= \overline{\Si}^{
IJKL})$$(63=28+35)$. 

It is well known  that the 70 real, physical 
scalars of ${\cal N}=8$ supergravity
parametrize the coset space $E_{7(+7)}/SU(8)$(even though $E_{7(+7)}$ 
symmetry is broken 
in the gauged theory) since 63 fields$(133-63=70)$ 
may be gauged
away by an $SU(8)$ rotation 
and can be represented by 
an element ${\cal V}(x)$ of the fundamental 56-dimensional representation
of $E_{7(+7)}$:
\bea
{\cal V}(x)=
 \mbox{exp}\left(
\begin{array}{ccc} \Lambda_{IJ}^{\;\;\;\;\;KL} 
& -\frac{1}{2\sqrt{2}}\;\phi_{IJPQ} \\
-\frac{1}{2\sqrt{2}}\;\overline{\phi}^{MNKL} &
\overline{\Lambda}^{MN}_{\;\;\;\;\;PQ}
\end{array}\right)= 
\left( \begin{array}{ccc} u_{ij}^{\;\;\;KL} 
& v_{ijPQ} \\
 \overline{v}^{mnKL} & \overline{u}^{mn}_{\;\;\;PQ}  \end{array} \right)
\nonu
\eea
where $SU(8)$ index pairs $[ij], \cdots$ and $SO(8)$ index 
pairs $[IJ], \cdots$
are antisymmetrized.
The 63 compact generators $\La$ can be
set to zero by fixing an $SU(8)$ gauge.
Moreover
$\phi_{IJKL}$ is a complex self-dual tensor describing the 35 
scalars $\bf 35_{v}$(the real part of $\phi_{IJKL}$)
and 35 pseudo-scalar fields $\bf 35_{c}$(the imaginary part of $\phi_{IJKL}$)
of ${\cal N}=8$ supergravity.
The maximally supersymmetric vacuum with $SO(8)$ symmetry, $\S^7$,
is where expectation values of both scalar and pseudo-scalar fields vanish.
Let us denote self-dual and anti-self-dual tensors of 
$SO(8)$ tensor as $C_{+}^{IJKL}$ and $C_{-}^{IJKL}$. Turning on the
scalar fields proportional to $C_{+}^{IJKL}$ yields an 
$SO(7)^{+}$-invariant vacuum. 
Likewise, turning on pseudo-scalar fields proportional to
$C_{-}^{IJKL}$ yields $SO(7)^{-}$-invariant vacuum. Both $SO(7)^{\pm}$
vacua are nonsupersymmetric. However, simultaneously turning on
both scalar and pseudo-scalar fields proportional to 
$C_{+}^{IJKL}$ and $C_{-}^{IJKL}$, respectively, one obtains $G_2$-invariant
vacuum with ${\cal N}=1$ supersymmetry \cite{dnw}. 
The most general vacuum expectation 
value of 56-bein preserving $G_2$-invariance can be parametrized by
\bea
\phi_{IJKL} = \frac{\la}{2\sqrt{2}} \left( \cos \al \;\; C_{+}^{IJKL}+
i \sin \al \;\; C_{-}^{IJKL} \right). 
\nonu
\eea
Then 
two scalars $\la$ and $\al$ fields in the $G_2$-invariant flow
parametrize a $G_2$-invariant subspace of the complete scalar manifold
$E_{7(+7)}/SU(8)$. The dependence of scalar 56-bein
\bea
{\cal V} (\la(x), \al(x)) = 
\left( \begin{array}{ccc} u_{ij}^{\;\;\;IJ} 
& v_{ijKL} \\
 \overline{v}^{\,klIJ} & \overline{u}^{\,kl}_{\;\;\;KL} \end{array} \right) 
\nonu
\eea
on $\la$ and $\al$ was obtained in \cite{dnw,ar2,aw} by exponentiating 
the vacuum expectation values $\phi_{IJKL}$ of $G_2$-singlet space.
We refer to the Appendix A in \cite{aw} for the 
explicit forms that one needs to know.

The structure of the scalar sector is encoded in 
the $SU(8)$-covariant $T$-tensor which is cubic in the 28-beins
and antisymmetric in the indices $[ij]$:
\bea
T_l^{\;kij} & = & 
\left(\overline{u}^{\,ij}_{\;\;IJ} +\overline{v}^{\,ijIJ} 
\right) \left( u_{lm}^{\;\;\;JK} 
\overline{u}^{\,km}_{\;\;\;KI}-v_{lmJK} \overline{v}^{\,kmKI} \right).
\nonu
\eea
The superpotential which leads to $G_2$-invariant flow is one of the 
eigenvalues of the symmetric tensor in $(ij)$:
\bea
A_1^{\;\;ij} & =& 
-\frac{4}{21} \,T_{m}^{\;\;ijm},
\nonu
\eea
which can be obtained by using some identities in $T$-tensor.
It turned out \cite{aw,dnw} 
that the $A_1^{\;\;ij}$ tensor has two distinct complex
eigenvalues, $z_1$ and $z_2$, with degeneracies 7 and 1 respectively 
to have the following form\footnote{It is easy to see that the eigenvalue 
$z_2$ in \cite{aw} becomes $z_1$ when we restrict to have $G_2$-invariant
flow by taking $\la' =\la$ and $\phi=\al$. Moreover 
we denote here $z_2$ by $z_3$ in \cite{aw} on the restricted subspace.
For the explicit form of $z_1$ in $A_1$ tensor, we refer to \cite{aw}.} 
\bea
A_1^{\;\;ij} = \mbox{diag} \left(z_1, z_1, z_1, z_1, z_1, z_1, z_1, 
z_2 \right)
\nonu
\eea
where the eigenvalues are functions of $\la$ and $\al$. In particular,
\bea 
z_2(\la, \al) & = & p^7+e^{7i\al} q^7 +7 \left( p^3 q^4 e^{4i\al} +
p^4 q^3 e^{3i\al}\right) 
\nonu
\eea
and we denote $p$, $q$ by the following hyperbolic functions of $\la$:
\bea
p = \cosh \left( \frac{\la}{2\sqrt{2}} \right), \qquad
q = \sinh \left( \frac{\la}{2\sqrt{2}} \right).
\nonu
\eea

\begin{table}
$$
\begin{array}{|c|c|c|c|c|}
\hline
$\mbox{Symmetry}$ 
& \la  & \al & V & W \nonu \\
\hline
   SO(8) & 0 & \mbox{any}  & - 6\,g^2 & 1 \nonu \\
\hline
 G_2 & \sqrt{2}\,\sinh^{-1}\!\sqrt{\frac{2}{5}(\sqrt{3}-1)} &
 \cos^{-1}\frac{1}{2}\sqrt{3-\sqrt{3}}
 & -\frac{216 \sqrt{2}}{25 \sqrt{5}}\,3^{1/4} g^2 
 & \sqrt{\frac{36 \sqrt{2}\,3^{1/4}}{25 \sqrt{5}}} \nonu \\
\hline
\end{array}
$$
\caption{\sl Summary of two supersymmetric critical points: symmetry group,
vacuum expectation values of fields, cosmological constants and 
superpotential.}
\end{table}

The superpotential is given by the eigenvalues of $A_1$ tensor and
the supergravity potential\footnote{The scalar potential can be written 
explicitly as 
\bea
V(\la, \al)= 2 g^2 \left( ( 7v^4 -7v^2 +3) \,c^3 s^4 +( 4v^2 -7) \,v^5 s^7 +
c^5 s^2 + 7 v^3 c^2 s^5 -3 c^3 \right)
\nonu
\eea
where $c = \cosh \left(\frac{\la}{\sqrt{2}} \right)$, 
$s = \sinh \left(\frac{\la}{\sqrt{2}} \right)$
and $v =\cos \al$
by getting all the components of $A_1$ and $A_2$ tensors \cite{aw,dnw}.}
can be written in terms of superpotential:
\bea
W(\la, \al) & = & |z_2|, \nonu \\
V(\la, \al) & = & g^2 \left[ \frac{16}{7} \left(\partial_{\la}
W \right)^2 + \frac{2}{7p^2 q^2}
\left(\partial_{\al}
W \right)^2  - 6 \,W^2 \right].
\label{spoten1}
\eea
The supersymmetric flow equations are \cite{aw}
\bea 
\partial_{r}\la & = & 
-\frac{8\sqrt{2}}{7}\,g \,\partial_{\la} W ,\nonu
\\
\partial_{r}\alpha & = & 
-\frac{\sqrt{2}}{7p^2 q^2}\,g \,\pa_{\al} W, \nonu \\ 
\partial_{r}A & = & \sqrt{2} \,g \,W.
\label{RGflow}
\eea

There exists a common supersymmetric critical point of both 
a scalar potential and a superpotential at 
$\sinh \left( \frac{\la}{\sqrt{2}} \right)= \sqrt{\frac{2}{5}(\sqrt{3}-1)}$,
$\cos \al = \frac{1}{2}\sqrt{3-\sqrt{3}}$ where 
$W=\sqrt{\frac{36 \sqrt{2}\,3^{1/4}}{25 \sqrt{5}}}$.
This implies the supersymmetry preserving vacua have negative
cosmological constant because the scalar potential $V$ at the 
critical points becomes $V=-6\,g^2 \,W^2$ due to the stationary 
of the superpotential $W$. 
This is the ${\cal N}=1$ supersymmetric critical point with $G_2$-symmetry.
The flow we are interested in is the one starting at the ${\cal N}=8$
supersymmetric point specified by $\la=0$ and ends at the non-trivial 
${\cal N}=1$ supersymmetric point. We summarize the properties of
these critical points in Table 1 and draw the 
scalar potential and superpotential by contour plots in Figure 1.

\begin{figure}
\begin{center}
\begin{minipage}[t]{6.8cm}
\centerline{\hbox{\psfig{file=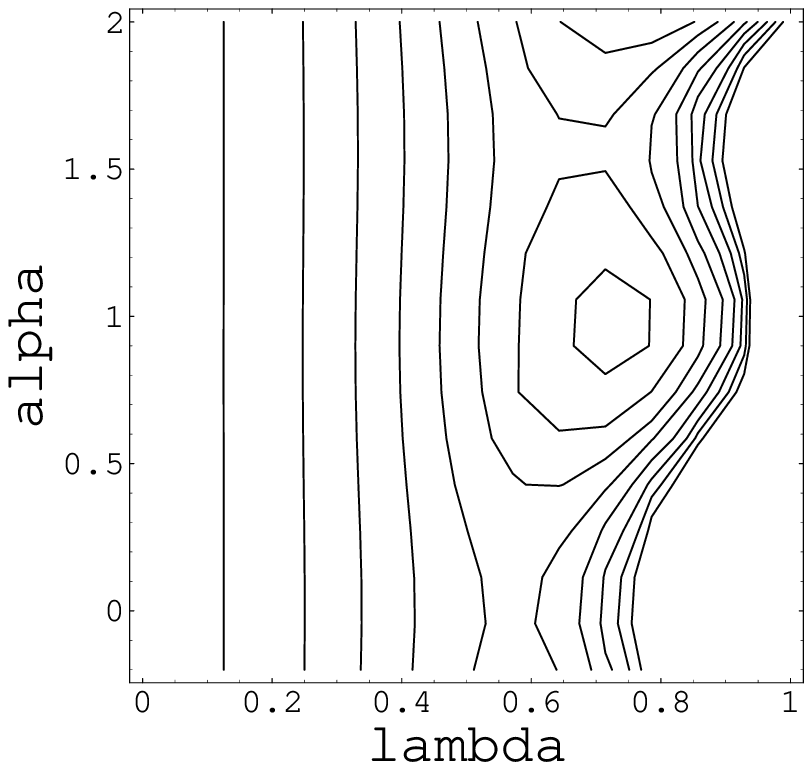,width=6.8cm}}}
\end{minipage}\hspace{1cm}
\begin{minipage}[t]{6.8cm}
\centerline{\hbox{\psfig{file=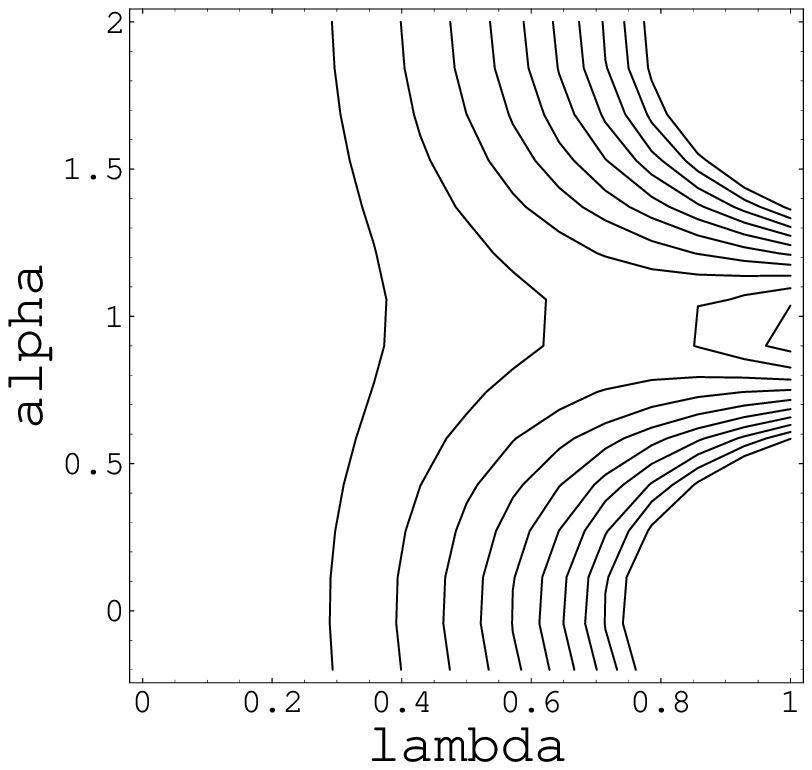,width=6.8cm}}}
\end{minipage}
\caption{\sl Contour plots of the scalar potential $V$(left)  and 
the superpotential $W$(right)  in the 4-dimensional gauged supergravity. 
The axes $(\lambda,\alpha)$ are two vevs that 
parametrize the $G_2$ invariant manifold in the 28-beins of the theory. 
The $SO(8)$ UV critical point is located at an arbitrary point on 
the line $\lambda=0$, whereas the $G_2$-invariant IR critical point 
is located at 
$(\lambda,\alpha)=\Bigl(\sqrt{2}\,\sinh^{-1}\!\sqrt{\frac{2}{5}(\sqrt{3}-1)},
\,\cos^{-1}\frac{1}{2}\sqrt{3-\sqrt{3}}\Bigr)\approx(0.732,0.973)$.
We have set the $SO(8)$ gauge coupling $g$ in the scalar potential as 
$g=1$. We also denote that  the $SO(7)^{+}$-invariant IR critical point 
is located at 
$(\lambda,\alpha)=\Bigl(\sqrt{2}\,\sinh^{-1}\!\sqrt{\frac{1}{2}(\frac{
3}{\sqrt{5}}-1)},
\,0 \Bigr)\approx(0.569,0)$ and 
the $SO(7)^{-}$-invariant IR critical point 
is located at 
$(\lambda,\alpha)=\Bigl(\sqrt{2}\,\sinh^{-1}\!\frac{1}{2},
\,\frac{\pi}{2} \Bigr)\approx(0.681,1.57)$. Note that $SO(7)^{\pm}$ critical
points of a scalar potential are not those of a superpotential, observed by 
the above contour plots.  }
\label{fig:spoten}
\end{center}
\end{figure}

\section{The $M$-theory lift of the $G_2$-invariant holographic RG flow}
\setcounter{equation}{0}
\setcounter{subsection}{0}

The $G_2$-invariant RG flow discussed in the previous section is a 
supersymmetric domain wall solution in the 4-dimensional gauged 
supergravity. 
The nonlinear metric ansatz of de Wit-Nicolai-Warner \cite{dnw} suggests 
that the RG flow solution in 4 dimensions can be lifted to a certain 
11-dimensional solution in $M$-theory. In this section, 
we carry out the $M$-theory lift of the $G_2$-invariant RG flow through 
a combinatoric use of the 4-dimensional RG flow equations and 
the 11-dimensional Einstein-Maxwell equations derived from the nonlinear 
metric ansatz of de Wit-Nicolai-Warner.

\subsection{The 11-dimensional metric for the $G_2$-invariant RG flow}

The consistency under the Kaluza-Klein compactification of 11-dimensional 
supergravity, or $M$-theory, requires that the 11-dimensional metric for 
the $G_2$-invariant RG flow is {\it not} simply a metric of product space 
{\it but} Eq.\ (\ref{metric}) for the warped product of $AdS_4$ space with 
a 7-dimensional compact manifold. 
Moreover, the 7-dimensional space becomes a warped-ellipsoidally deformed 
${\bf S}^7$ and its metric is uniquely determined through the nonlinear 
metric ansatz developed in \cite{dnw,dn87}. 
The 7-dimensional inverse metric is given by the formula 
(\ref{internalmetric}):
\begin{equation}
g^{mn}=\frac{1}{2}\,\Delta
\left[\Kop^{mIJ}\Kop^{nKL}+(m \leftrightarrow n)\right]
\left(u_{ij}^{~~IJ}+v_{ijIJ}\right)
\left(\overline{u}^{\,ij}_{~~KL}+\overline{v}^{\,ijKL}\right),
\label{7dmet}
\end{equation}
where $\Kop^{mIJ}$ denotes the Killing vector on the round ${\bf S}^7$ 
with 7-dimensional coordinate indices $m, n=5,\dots,11$ as well as $SO(8)$ 
vector indices $I, J=1,\dots,8$. The $u_{ij}^{~~IJ}$ and $v_{ijIJ}$ are  
28-beins in 4-dimensional gauged supergravity and are parametrized 
by the $AdS_4$ vacuum expectation values(vevs), $\lambda$ and $\alpha$, 
associated with the spontaneous compactification of 11-dimensional 
supergravity. 

The 28-beins $(u, v)$ or two vevs $(\lambda, \alpha)$ are given by functions 
of the $AdS_4$ radial coordinate $r=x^4$. The metric formula (\ref{7dmet}) 
generates the 7-dimensional metric from the two input data of $AdS_4$ vevs 
$(\lambda,\alpha)$. The $G_2$-invariant RG flow subject to (\ref{RGflow}) 
is a trajectory in $(\lambda,\alpha)$-plane and is parametrized by the 
$AdS_4$ radial coordinate. 
Hereafter, instead of $(\lambda,\alpha)$, we will use $(a,b)$ defined by
\begin{eqnarray}
a &\equiv& \cosh\!\left(\frac{\lambda}{\sqrt{2}}\right)
+\cos\alpha\,\sinh\!\left(\frac{\lambda}{\sqrt{2}}\right),\nonumber\\
b &\equiv& \cosh\!\left(\frac{\lambda}{\sqrt{2}}\right)
-\cos\alpha\,\sinh\!\left(\frac{\lambda}{\sqrt{2}}\right).
\label{abvevs}
\end{eqnarray}

Let us first introduce the standard metric of a 7-dimensional ellipsoid. 
Using the diagonal matrix $Q_{AB}$ given by\footnote{
Note that our assignment of $(a,b)$ in $Q_{AB}$ is different from (5.46) 
in \cite{dnw}. The former is obtained by switching $a$ and $b$ and 
by reparametrizing the ${\bf R}^8$ coordinates $X^A$ in the latter.}
$$
Q_{AB}={\rm diag}\left(1,\dots,1,\frac{a^2}{b^2}\right),
$$
the metric of a 7-dimensional ellipsoid with the eccentricity 
$\sqrt{1-b^2/a^2}$ can be written as
\begin{equation}
ds_{EL(7)}^2(a,b) =dX^A Q^{-1}_{AB}\,dX^B 
-\frac{b^2}{\xi^2}\left(X^A \delta_{AB}\,dX^B\right)^2, \label{7dmet2}
\end{equation}
where the ${\bf R}^8$ coordinates $X^A (A=1,\dots,8)$\footnote{
The ${\bf R}^8$ coordinates $X^A$ are given in terms of 
the 7-dimensional coordinates $y^m$ by
\begin{eqnarray*}
&& X^1 = s_1 \,s_2 \,s_3 \,s_4 \,s_5 \,s_6 \,s_7, \quad
X^2 = c_1 \,s_2 \,s_3 \,s_4 \,s_5 \,s_6 \,s_7, \quad
X^3 = c_2 \,s_3 \,s_4 \,s_5 \,s_6 \,s_7, \quad
X^4 = c_3 \,s_4 \,s_5 \,s_6 \,s_7, \\ &&
X^5 = c_4 \,s_5 \,s_6 \,s_7, \quad
X^6 = c_5 \,s_6 \,s_7, \quad
X^7 = c_6 \,s_7, \quad
X^8 = c_7,
\end{eqnarray*}
where $s_m \equiv \sin y^m$ and $c_m \equiv \cos y^m (m=1,\dots,7)$.}
are constrained on the unit round ${\bf S}^7$, $\sum_{A}(X^A)^2 =1$, 
and $\xi^2 =b^2 \,X^A Q_{AB} X^B$ is a quadratic form on the ellipsoid. 
The standard metric (\ref{7dmet2}) can be rewritten in terms of the 
7-dimensional coordinates $y^m$ such that
\begin{equation}
ds_{EL(7)}^2(a,b)=a^{-2}\,\xi^2\,d\theta^2 
+\sin^2 \theta\,d\Omega_6^2, \label{7dmet3}
\end{equation}
where $\theta =y^7$ is the fifth coordinate in 11 dimensions 
and the quadratic form $\xi^2$ is now given by
\bea
\xi^2 =a^2 \cos^2 \theta + b^2 \sin^2 \theta,
\label{xi}
\eea
which turns to 1 for the round ${\bf S}^7$ with $(a,b)=(1,1)$. 
We will see that the 7-dimensional metric generated from the formula 
(\ref{7dmet}) is a ``warped'' metric of a 7-dimensional ellipsoid 
(\ref{7dmet2}), where 
the geometric parameters $(a,b)$ for the 7-ellipsoid can be identified 
with the two vevs $(a,b)$ defined in (\ref{abvevs}). 
It is the reason why we have introduced $(a,b)$ in (\ref{abvevs}) and 
prefer them rather than the original vevs $(\lambda,\alpha)$. 
In fact the Ricci tensor components generated from the 11-dimensional metric 
which we will derive in this subsection become much simpler in terms of 
$(a,b)$ (See Appendix B). 

To calculate the 7-dimensional metric by using the formula (\ref{7dmet}), 
first we have to specify the Killing vector on the round ${\bf S}^7$. 
It is given by
\begin{equation}
\Kop_m^{~~IJ}=(\Gamma^{IJ})_{AB}\,L^2 
\left(X^A \partial_m X^B -X^B \partial_m X^A\right), \label{killing}
\end{equation}
where $X^A (A=1,\dots,8)$ are the ${\bf R}^8$ coordinates of the unit round 
${\bf S}^7$ and parametrized by the 7-dimensional coordinates $y^m$. 
The $8\times8$ matrices $\Gamma^{IJ}$ are the $SO(8)$ generators\footnote{
The $SO(8)$ matrices $\Gamma^{IJ}$ are defined by 
$\Gamma^{IJ}=\frac{1}{2}[\Gamma^I,\Gamma^J]$, $\Gamma^{I1}=\Gamma^I$, 
where $\Gamma^I (I=2,\dots,8)$ are the $SO(7)$ gamma matrices. 
We refer to Appendix B in \cite{aw2} for further details about 
these matrices.}
which rotate the ${\bf R}^8$ indices $A,B$ into the $SO(8)$ indices $I,J$. 

Applying the Killing vector (\ref{killing}) together with 
the 28-beins to the metric formula (\ref{7dmet}), we obtain a ``raw'' 
inverse metric $g^{mn}$ including the warp factor $\Delta$ not yet determined. 
Substitution of the raw inverse metric into the definition
$$
\Delta^{-1}\equiv\sqrt{\det(g^{mn}\,\Gop_{\,np})},
$$
which is equivalent to (\ref{warp}), provides a self-consistent equation 
for $\Delta$. 
For the $G_2$-invariant RG flow, solving this equation yields the warp factor
$$
\Delta(x, y)=a^{-1}\,\xi^{-{4 \over 3}}
$$
where $\xi$ is given in (\ref{xi}). Then we substitute this warp factor 
into the raw inverse metric to obtain the 7-dimensional metric: 
\begin{equation}
ds_7^2 =g_{mn}(y) \,dy^m dy^n
=(\Delta\,a)^{1 \over 2}L^2 \left(a^{-2}\,\xi^2\,d\theta^2 
+\sin^2 \theta\,d\Omega_6^2\right), \label{ellip}
\end{equation}
where one can see that the standard 7-dimensional metric (\ref{7dmet3}) 
is warped by a factor $(\Delta\,a)^{1 \over 2}$.
The nonlinear metric ansatz finally combines the 7-dimensional metric 
(\ref{ellip}) with the four dimensional metric with warp factor to yield
the warped 11-dimensional metric: 
\begin{equation}
ds_{11}^2 =\Delta^{-1}\left(dr^2 +e^{2 A(r)}
\eta_{\mu\nu}dx^\mu dx^\nu \right)+ds_7^2,
\label{11dmet}
\end{equation}
where $r=x^4$ and $\mu,\nu=1,2,3$ with $\eta_{\mu\nu}={\rm diag}(-,+,+)$. 

Note that two vevs $(\lambda,\alpha)$ as well as the domain wall amplitude 
$A(r)$ have non-trivial $r$-dependence subject to the RG flow equations 
(\ref{RGflow}) so that the warp factor and the 7-dimensional 
metric also depend on $r$ through the vevs. As we see in (\ref{RGflow}), 
$\partial_r \lambda$ and $\partial_r \alpha$ are proportional to 
the $\lambda$-, $\alpha$-derivatives of $AdS_4$ superpotential, respectively. 
We also see that as a characteristic feature of supersymmetric RG flows 
both $SO(8)$ and $G_2$ invariant critical points are {\it simultaneous} 
stationary 
points of the scalar potential and the superpotential
in the 4-dimensional gauged supergravity(See Figure \ref{fig:spoten}). 
Therefore $(\partial_r \lambda,\partial_r \alpha)=(0,0)$ at both $SO(8)$ and 
$G_2$ invariant critical points and the scalar potential is related to the 
superpotential by $V=-6\,g^2\,W^2$ as shown in (\ref{spoten1}) . 
On the other hand, $\partial_r A$ is proportional to the 
superpotential itself and is relevant in both critical points. 
Thus one can say that the 11-dimensional geometries of critical point 
theories are completely determined by both  vevs $(a,b)$ and  scalar 
potential $V$ without knowing the RG flow equations, that is, 
the $r$-dependence of the vevs. 

Furthermore, the superkink amplitude $A(r)$ asymptotically behaves as 
$A(r)=r/r_{\rm cr}$ near the critical points so that one can read 
$1/r_{\rm cr} \propto \dsqrt{-V_{\rm cr}}$ via the last equation 
in (\ref{RGflow}). 
Provided that the two vevs and the slope of the superkink 
$r_{\rm cr}$(or the value of scalar potential $V_{\rm cr}$) are known 
at each critical point, the 11-dimensional metric (\ref{11dmet}) reproduces 
the metric derived in \cite{dnw}. 
The $SO(8)$-invariant critical point is specified by $(a,b)=(1,1)$ and 
$A(r)=2 r/L$ with the $AdS_4$ radius $r_{\rm cr}=L/2$. 
The metric (\ref{11dmet}) 
is so normalized for the maximally symmetric case as to generate the Ricci 
tensor:
$$
R_{M}^{~~N}=\frac{6}{L^2}\,{\rm diag}\left(-2,-2,-2,-2,~1,~\dots,~1\right),
$$
which fixes the round ${\bf S}^7$ radius to be $L$, 
twice the $AdS_4$ radius, as expected. 
Since the superpotential satisfies $W=1$ at the $SO(8)$-invariant 
critical point, this normalization fixes the $SO(8)$ gauge coupling $g$ 
to be $\sqrt{2}\,L^{-1}$.

\subsection{The Einstein-Maxwell equations for the $G_2$-invariant RG flow}

The non-trivial $r$-dependence of vevs requires that the 11-dimensional 
Einstein-Maxwell equations become consistent {\it not only} at critical 
points {\it but also} along the supersymmetric RG flow connecting 
the two critical points. 
Even though Ref.\ \cite{dn87} provides another formula involving both 
the metric and the 4-form gauge field strength, the analysis in mid eighties 
before AdS/CFT correspondence just focused on the critical point theories 
with constant vevs so that the $G_2$-invariant field strength ansatz 
in \cite{dnw} is not quite general: they hold only for constant vevs. 
For solutions with varying scalars, the ansatz for the field strength 
will be a little more complicated.
Although the results of \cite{dn87} provide both 
the metric and the 4-form gauge field strength for nonconstant scalars,
it is not very convenient to use. It is an open problem to identify our
results here with those in \cite{dn87}. 
In this subsection, we will apply the {\it most general} 
$G_2$-invariant ansatz for an 11-dimensional 3-form gauge field by 
acquiring the 
$r$-dependence of vevs and will derive the 11-dimensional Einstein-Maxwell 
equations corresponding to the $G_2$-invariant RG flow. Hereafter we use 
Greek indices for the 3-dimensional spacetime of a membrane world volume, 
whereas Latin indices for the $G_2$-invariant 6-dimensional unit sphere. 
Specifically, $\mu,\nu,\dots =1,2,3$ and $m,n,\dots =6,\dots,11$.

As a natural extension of the Freund-Rubin parametrization \cite{fr}, 
the 3-form gauge field with 3-dimensional membrane indices may be defined 
by \cite{cpw,jlp}
\begin{equation}
A_{\mu\nu\rho} = -e^{3A(r)}\,\tW(r,\theta)\,
\epsilon_{\mu\nu\rho},\label{ansatz1}
\end{equation}
where $\tW(r, \theta)$ is a ``geometric'' superpotential \cite{kw} 
which will be relevant in search of a membrane probe moduli 
space(for example, \cite{jlp}) and to be determined. 
This is simply geometric superpotential 
$\tW(r, \theta)$ times the volume form on the membrane measured using 
the four-dimensional metric (\ref{domain}). 
The exponential factor $e^{3A(r)}$ will be compensated by 
the same factor arising from the 11-dimensional metric (\ref{11dmet}) 
when we derive the geometric superpotential.
The $\theta$-dependence of $\tW(r, \theta)$ is {\it essential} 
to achieve the $M$-theory lift of the RG flow.

The other components of 3-form gauge field have the indices on the round 
${\bf S}^6$ which is isomorphic to the coset space $G_2/SU(3)$ including 
the $G_2$-symmetry of our interest \cite{gw}. 
As is explicitly performed in \cite{gw}, one can geometrically construct 
the $G_2$-covariant tensors living on the round ${\bf S}^6$ by using the 
imaginary octonion basis of ${\bf S}^6$. 
Since one cannot construct any $G_2$-invariant vector through contracting 
those tensors, the $G_2$-covariance of the gauge field is achieved 
by using the $G_2$-covariant tensors only. Thus we arrive at the 
most general $G_2$-invariant ansatz:
\begin{eqnarray}
A_{4mn} &=& g(r,\theta)\,F_{mn}, \nonumber\\
A_{5mn} &=& h(r,\theta)\,F_{mn}, \nonumber\\
A_{mnp} &=& h_1 (r,\theta)\,T_{mnp} +h_2 (r,\theta)\,S_{mnp},\label{ansatz2}
\end{eqnarray}
where $m,n,p$ are the ${\bf S}^6$ indices and run from 6 to 11. 
The almost complex structure on the ${\bf S}^6$ is denoted by $F_{mn}$ 
and obeys $F_{mn}F^{\,nl}=-\delta_m^{\,l}$. 
The $S_{mnp}$ is the parallelizing torsion of the unit round ${\bf S}^7$ 
projected onto the ${\bf S}^6$, 
while the $T_{mnp}$ denotes the 6-dimensional Hodge dual of $S_{mnp}$.
We refer to \cite{gw} for further details about the tensors. 
In the ansatz (\ref{ansatz2}) in addition to the $r$-dependence of 
coefficients, the $A_{4mn}$ is new compared with the ansatz in previous works 
\cite{dnw,gw}(See for example (3.4) in \cite{gw}). 
The above ansatz for gauge field is in fact the most general one which 
preserves the $G_2$-invariance and is consistent with the 11-dimensional 
metric (\ref{11dmet}).\footnote{In general, one cannot neglect $A_{\mu mn}$, 
$A_{\mu\nu 4}$, $A_{\mu\nu 5}$ and $A_{\mu 45}$ by the $G_2$-invariance only. 
However, except for $A_{\mu 45}$, such gauge fields provide the field 
strengths $F_{\mu mnp}$, $F_{\mu mn4}$, $F_{\mu mn 5}$ and $F_{\mu\nu45}$ 
which generate off-diagonal $(\mu m)$-, $(\mu 4)$-, $(\mu 5)$-components 
in the RHS of the Einstein equation. 
The Ricci tensor in the LHS, however, do not contain such off-diagonal 
elements and forbids the $A_{\mu mn}$, $A_{\mu\nu 4}$ and $A_{\mu\nu 5}$ 
to arise in the gauge field ansatz. The $A_{\mu 45}$ depends on 
$(r,\theta)$ only so that it does not affect field strengths at all. 
It can be zero as a gauge fixing.}

Through the definition $F_{MNPQ}\equiv 4\,\partial_{[M}A_{NPQ]}$, 
the ansatz (\ref{ansatz1}) generates the field strengths
\begin{eqnarray}
F_{\mu\nu\rho 4} &=& e^{3A(r)}
\,W_r (r,\theta)\,\epsilon_{\mu\nu\rho},\nonumber\\
F_{\mu\nu\rho 5} &=& e^{3A(r)}
\,W_{\theta}(r,\theta)\,\epsilon_{\mu\nu\rho},\label{fst1}
\end{eqnarray}
while the ansatz (\ref{ansatz2}) provides\footnote{
Our convention of epsilon tensors is that the tensors with lower 
indices are purely numerical. The 6-dimensional epsilon tensor in 
(\ref{fst2}) is the only exception and is defined as a constant tensor 
density on the round ${\bf S}^6$, that is
$\eps6_{mnpqrs}=\dsqrt{g_6}\, \epsilon_{mnpqrs}$,
where $g_6$ is the determinant of the ${\bf S}^6$ metric.}
\begin{eqnarray}
F_{mnpq} &=& 2\,h_2 (r,\theta)\,\eps6_{mnpqrs} F^{rs}, \nonumber\\
F_{5mnp} &=& \th_1 (r,\theta)\,T_{mnp}
+\th_2 (r,\theta)\,S_{mnp}, \nonumber\\
F_{4mnp} &=& \th_3 (r,\theta)\,T_{mnp}
+\th_4 (r,\theta)\,S_{mnp}, \nonumber\\
F_{45mn} &=& \th_5 (r,\theta)\,F_{mn}, \label{fst2}
\end{eqnarray}
where the coefficient functions which depend on both $r$ and $\theta$
are given by
\begin{eqnarray}
W_r &=& e^{-3A}\partial_r (e^{3A}\tW),\qquad 
W_\theta ~=~ e^{-3A}\partial_\theta (e^{3A}\tW),\nonumber\\
\th_1 &=& \partial_\theta h_1 -3h,\qquad
\th_2 ~=~ \partial_\theta h_2, \nonumber\\
\th_3 &=& \partial_r h_1 -3g,\qquad \th_4 ~=~ \partial_r h_2, \nonumber\\
\th_5 &=& \partial_r h -\partial_\theta g. \label{fst3}
\end{eqnarray}
Comparing with the ansatz in the previous works \cite{dnw,gw}, the mixed 
field strengths $F_{\mu\nu\rho5}$, $F_{4mnp}$ and $F_{45mn}$ are new. 
In fact, they are not forbidden to arise by the $G_2$-invariance 
once we suppose that the 4-dimensional metric has the domain wall factor 
$e^{3A(r)}$ which breaks the 4-dimensional conformal invariance. 
At both $SO(8)$-invariant UV and $G_2$-invariant IR critical points, 
the 4-dimensional spacetime becomes asymptotically $AdS_4$ and the mixed 
field strengths should vanish there. 
Especially, the $F_{\mu\nu\rho5}$ and $F_{45mn}$ are proportional to 
$W_\theta$ and $\th_5$, respectively, so that they must be subject to 
the non-trivial boundary conditions:
\begin{equation}
W_\theta =0, \quad \th_5=0,
\label{bc}
\end{equation}
at both UV and IR critical points. 
It is easy to see that the $F_{4mnp}$ goes to zero 
at both critical points without imposing any boundary condition 
since the field strength comes from the $r$-derivative of 
$A_{mnp}$(note that $r=x^4$) and 
the vevs become $r$-independent at both critical points.\footnote{
We implicitly suppose that the field strength functions $\th_1,\dots,\th_5$ 
are finally written as polynomials of vevs $(a,b)$ after the 11-dimensional 
Einstein-Maxwell equations are solved by imposing the RG flow equations.
In Section 3.3, we will see that  is the case.} 

We also have the 11-dimensional Bianchi identity 
$\partial_{[M}F_{NPQR]}\equiv0$ which reads
\begin{equation}
\partial_\theta W_r -\partial_r W_\theta = 3\,(\partial_r A)\,W_\theta,\quad
\partial_r (\th_1 +3h) = \partial_\theta (\th_3 +3g),\quad 
\partial_r \th_2 = \partial_\theta \th_4,
\label{bian}
\end{equation}
and simply provides the integrability conditions for the gauge fields 
$e^{3A}\tW$, $h_1$ and $h_2$. The $W_r$ and $W_\theta$ obtained by solving 
the 11-dimensional Einstein-Maxwell equations should satisfy the first 
equation by imposing the RG flow equations.

Applying the field strength ansatz (\ref{fst1}), (\ref{fst2}) and the metric 
(\ref{11dmet}) to the 11-dimensional Maxwell equation in 
(\ref{fieldequations}), we 
obtain(See (\ref{M1}-\ref{M4}) in Appendix A also)
\begin{eqnarray}
a^3 D_\theta \th_1 +L^2 D_r \th_3 &=& 2L\,\xi^{-2}
\left(W_r \th_2 -W_\theta \th_4 \right), \nonumber\\
a^3 D_\theta \th_2 +L^2 D_r \th_4 &=& 2L\,\xi^{-2}
\left(W_\theta \th_3 -W_r \th_1 \right) 
+\frac{12\,a\,\xi^2\,h_2}{\sin^2 \theta}, \nonumber\\
a^3 \tD_\theta W_\theta +L^2 \tD_r W_r &=& 
-\frac{8\,a^3\,\xi^4}{L^5 \sin^6 \theta}
\left(\th_1 \th_4 -\th_3 \th_2 -3\,h_2 \th_5 \right), \label{maxf1}
\end{eqnarray}
with $a^3 \th_1$ and $L^2 \th_3$ solved to be
\begin{eqnarray}
a^3 \th_1 &=& 2L\,\xi^{-2}\,W_r h_2 
-\frac{1}{4}\,L^2 a^2 \xi^{-2} \sin^2 \theta\, {\cal D}_r \th_5, \nonumber\\
L^2 \th_3 &=& -2L\,\xi^{-2}\,W_\theta h_2 
+\frac{1}{4}\,L^2 a^2 \xi^{-2} \sin^2 \theta\, {\cal D}_\theta \th_5.
\label{maxf2}
\end{eqnarray}
Here we have defined
\begin{eqnarray*}
(D_\theta, D_r) &\equiv& \hat{e}^{-1}\left(\partial_\theta,
\partial_r \right)\hat{e},
\qquad \hat{e} ~\equiv~ \xi^2 e^{3A},\nonumber\\
(\tD_\theta, \tD_r) &\equiv& \tilde{e}^{-1}\left(\partial_\theta,
\partial_r \right)\tilde{e},
\qquad \tilde{e} ~\equiv~ a^{-3}\,\xi^{-4}\sin^6 \theta,\nonumber\\
({\cal D}_\theta, {\cal D}_r) &\equiv& \vep^{-1}\left(\partial_\theta,
\partial_r \right) \vep,
\qquad \vep ~\equiv~ a^2 e^{3A} \sin^2 \theta.
\end{eqnarray*}

The first equation in (\ref{maxf1}) is trivially satisfied 
via substitution of (\ref{maxf2}). 
The second equation becomes\footnote{
We use the notation $h_2^{'}\equiv\partial_\theta h_2$ and 
$\dot{h}_2\equiv\partial_r h_2$.}
\begin{eqnarray}
&& a^3 D_\theta h_2^{'} +L^2 D_r \dot{h}_2
+a^3 \left[4\,f^2 L^2\,a^{-6}\,\xi^{-4}
-\frac{12\,a^{-2}\,\xi^2}{\sin^2 \theta}\right] h_2 \nonumber\\
&& \qquad\qquad\qquad = \frac{1}{2}\,L\,a^{-1} \xi^{-4} \sin^2 \theta 
\left(a^3 \,W_\theta {\cal D}_\theta +L^2 \,W_r {\cal D}_r \right)\th_5,
\label{1stmax}
\end{eqnarray}
where the parameter $f$ defined by
\begin{equation}
f \equiv \frac{\dsqrt{a^3 \,W_\theta^2 +L^2 \,W_r^2}}{L} \label{freund}
\end{equation}
corresponds to the Freund-Rubin parameter in each critical point theory.
Note that (\ref{1stmax}) goes to (4.7) in \cite{dnw} at both UV and IR 
critical points. To see that, one only has to remember that $\th_5 =0$ 
and $h_2$ becomes $r$-independent at both critical points. 
Then one can see that, in (\ref{1stmax}), the RHS and the second term of 
the LHS vanish at both critical points to reduce the equation to 
(4.7) in \cite{dnw}. To be more precise, one also has to show that 
the generalized Freund-Rubin parameter $f$ becomes constant at both 
critical points. 
This can be seen from the boundary condition $W_\theta =0$ and the 
11-dimensional Einstein equation discussed below which certifies that 
the $W_r$ becomes constant at the critical points. 

The third equation becomes
\begin{eqnarray}
a^3 \tD_\theta W_\theta +L^2 \tD_r W_r &=&
-\frac{16\,\xi^2}{L^6 \sin^6 \theta}
\biggl[\, a^3 \,W_\theta \,h_2^{'}\, h_2
+L^2 \,W_r \,\dot{h}_2\, h_2 
-\frac{3}{2}\,L\,a^3 \xi^2 \,h_2 \th_5 \nonumber\\
&& -\frac{1}{8}\,L\,a^2 \sin^2 \theta
\left(a^3 \,h_2^{'}\, {\cal D}_\theta 
+L^2 \,\dot{h}_2\, {\cal D}_r \right) \th_5 \biggr],
\label{2ndmax}
\end{eqnarray}
which is new and is relevant only at the intermediate points of RG flow 
connecting the two critical points. One can easily see that the equation is 
satisfied at the two critical points(or end points of RG flow) 
by recalling the boundary condition (\ref{bc}) and that both $h_2$ and 
$W_r$ become $r$-independent as well as two vevs $(a,b)$ there. 

To summarize, the 11-dimensional Maxwell equations (\ref{maxf1}) reduce to 
the two independent equations (\ref{1stmax}) and (\ref{2ndmax}) 
which consist of the $G_2$-invariant 11-dimensional field equations 
together with the 11-dimensional Einstein equation discussed below. 
They should be satisfied {\it everywhere along} the supersymmetric 
RG flow if the $M$-theory solution corresponding to the $G_2$-invariant RG 
flow exists. In other words, the equations must be consistent with each other 
{\it provided} that the $r$-dependence of field strengths is only through 
the vevs which are subject to the RG flow equations (\ref{RGflow}). 

Via the field strengths ansatz (\ref{fst1}), (\ref{fst2}) and 
the warped metric (\ref{11dmet}), the 11-dimensional Einstein equation 
in (\ref{fieldequations}) goes to(See (\ref{E1}-\ref{E5}) 
in Appendix A also)\footnote{
Other nonzero components are
$$
R_2^{~2}=R_3^{~3}=R_1^{~1}, \quad 
R_7^{~7}=R_8^{~8}=R_{9}^{~9}=R_{10}^{~10}=R_{11}^{~11}=R_6^{~6}, \quad 
R_4^{~5}=L^{-2}\,a^{3} \,R_5^{~4}.
$$}
\begin{eqnarray}
R_1^{~1} &=& -\frac{4}{3}\,c_3\, f^2
-\frac{1}{3}\,c_1 \left[H +\frac{12\,a\,\xi^2\,h_2^2}{\sin^2 \theta}
+c_2\,\th_5^2\right],
\nonumber\\
R_4^{~4} &=& R_1^{~1} +2\,c_3\,L^{-2}\,a^3\,W_\theta^2 
+c_1 \left[L^2 \left(\th_3^2 +\th_4^2\right)+c_2\,\th_5^2\right],
\nonumber\\
R_5^{~5} &=& R_1^{~1} +2\,c_3\, W_r^2 
+c_1 \left[a^3 \left(\th_1^2 +\th_2^2\right)+c_2\,\th_5^2\right],
\nonumber\\ 
R_6^{~6} &=& R_1^{~1} +2\,c_3\,f^2 
+\frac{1}{2}\,c_1 \left[H +\frac{16\,a\,\xi^2\,h_2^2}{\sin^2 \theta}
+\frac{2}{3}\,c_2\,\th_5^2 \right],
\nonumber\\
R_5^{~4} &=& -2\,c_3\, W_r W_\theta 
+c_1\,L^2 \left(\th_1 \th_3 +\th_2 \th_4\right), \label{einst}
\end{eqnarray}
where we have defined
$$
c_1 \equiv \frac{8\,a^{-1}\,\xi^{2 \over 3}}{L^8 \sin^6 \theta},\qquad
c_2 \equiv \frac{3}{4}\,L^2\,a^2\,\xi^{-2} \sin^2 \theta, \qquad
c_3 \equiv a^{-4}\,\xi^{-{16 \over 3}},
$$
and
\begin{equation}
H \equiv a^3 \left(\th_1^2 +\th_2^2\right)+L^2 \left(\th_3^2 +\th_4^2\right).
\label{Hdef}
\end{equation}
The Ricci tensor components in the LHS are generated by the warped 
11-dimensional metric (\ref{11dmet}) as listed in Appendix B. 
Note that the off-diagonal (4,5), (5,4)-components arise due to $\th_5$, 
the $\theta$-dependence of $\tW(r, \theta)$ and the $r$-dependence of $h_2$. 
The boundary conditions (\ref{bc}) ensure 
that those off-diagonal components vanish and the diagonal $R_4^{~4}$ and 
$R_5^{~5}$ components degenerate to $R_1^{~1}$ and $R_6^{~6}$, respectively, 
at both critical points \cite{dnw}. 

Even though one does not know an appropriate ansatz for $h_2$ and $\th_5$, 
the Einstein equations (\ref{einst}) can be solved with respect to 
certain combinations of field strength squares. 
For instance, the generalized Freund-Rubin parameter $f$ can be solved to be
\begin{eqnarray}
f = \dsqrt{\frac{R_4^{~4}+R_5^{~5}+4R_1^{~1}+6R_6^{~6}}{-2\,c_3}}. \label{f2}
\end{eqnarray}
One can also see that $h_2$, $\th_5$ and $H$ obey the following equations
\begin{eqnarray}
\left(\frac{12\,a\,\xi^2}{\sin^2 \theta}\right)h_2^2 -c_2\,\th_5^2
&=& \frac{3}{c_1}\left(R_1^{~1}+2R_6^{~6}\right),\label{h2h5}\\
H+2\,c_2\,\th_5^2 &=& \frac{2}{c_1}
\left(R_1^{~1}+R_4^{~4}+R_5^{~5}+3R_6^{~6}\right).\label{Hh5}
\end{eqnarray}
In order to determine all field strengths $W_r$, $W_\theta$, $h_2$ and 
$\th_5$, separately, one has to specify a certain functional form of 
$\th_5$ as an ansatz. 
Substitution of this ansatz into the Maxwell equation (\ref{1stmax}) 
may uniquely determine $\th_5$. The other field strengths are 
automatically generated by substituting the $\th_5$ into other 
11-dimensional field equations and by invoking the RG flow equations. 
Suppose that there exists such a non-trivial solution to the $\th_5$, 
there arises a non-trivial question whether the obtained field strengths 
are consistent with each other {\it everywhere along} the RG flow. 
It is the subject in Section 3.3 to answer this question.

\subsection{The $M$-theory lift of the $G_2$-invariant RG flow: 
An exact solution in 11-dimensions}

In Section 3.2 we have derived the 11-dimensional 
Einstein-Maxwell equations from the warped metric (\ref{11dmet}) 
and the field strength ansatz (\ref{fst1}) and (\ref{fst2}). 
The 11-dimensional field equations are closed within the field strengths 
$W_r$, $W_\theta$, $h_2$ and $\th_5$ although they cannot be solved 
separately without imposing certain ansatz for them. 
Our final goal is to determine all the field strengths as polynomials of 
vevs $(a,b)$ by imposing their $r$-dependence controlled by the RG flow 
equations (\ref{RGflow}) and to confirm whether those field strengths 
satisfy the Maxwell equations (\ref{1stmax}) and (\ref{2ndmax}) 
{\it everywhere along} the $G_2$-invariant RG flow. 
If this is the case, the field strengths $(W_r,W_\theta,h_2,\th_5)$ and 
the 11-dimensional metric (\ref{11dmet}) consist of an exact $M$-theory 
solution {\it provided} that the $r$-dependence of vevs $(a,b)$ and 
the superkink amplitude $A(r)$ are subject to the 4-dimensional RG flow 
equations (\ref{RGflow}). 

The key idea of the nonlinear metric ansatz in \cite{dnw} is to recognize
that a certain compactification of 11-dimensional supergravity
goes to a critical point of $d=4$, ${\cal N}=8$ gauged supergravity
{\it after} a truncation of the full 11-dimensional mass spectrum to
its massless sector(including the scalar fields in 28-beins).
For the supersymmetric critical points, such a truncation becomes consistent
only if the supersymmetry transformation can be closed within the massless 
sector by imposing the nonlinear metric ansatz.
This means that the metric formula (\ref{internalmetric}) derived from
the nonlinear metric ansatz is valid even for the scalar fields {\it varying}
along the $AdS_4$ radial coordinates.
On the other hand, the $G_2$-invariant RG flow is a supersymmetric
domain wall solution with varying scalars interpolating two critical
points with constant scalar vevs.
With the aforementioned in mind, we have enough reason to believe the 
existence of an $M$-theory lift of the ${\cal N}=1$, $G_2$-invariant RG flow.

As we see in the 11-dimensional metric (\ref{11dmet}), 
the two vevs $(\lambda,\alpha)$ arise in the metric only through the 
combinations of $(a,b)$ given in (\ref{abvevs}). 
The Ricci tensor calculated from the metric therefore contains $(a,b,A)$
and their $r$-derivatives only(See Appendix B). 
We have shown in Section 3.1 that the $(a,b)$ are in fact the geometric 
parameters which specify the warped ellipsoidal deformation of ${\bf S}^7$ 
through $Q_{AB}$. With taking these facts into account, we prefer 
to use $(a,b)$ and their flow equations rather than $(\lambda,\alpha)$ 
and the original flow equations (\ref{RGflow}). 
In terms of $(a,b)$ the flow equations 
(\ref{RGflow}) read in symmetric form
\begin{eqnarray}
\partial_r a &=& -\frac{8}{7L}
\biggl[a^2\,\partial_a W(a,b) +(ab-2)\,\partial_{\,b} W(a,b)\biggr], 
\nonumber\\
\partial_r b &=& -\frac{8}{7L}
\biggl[b^2\,\partial_{\,b} W(a,b) +(ab-2)\,\partial_a W(a,b)\biggr],
\nonumber\\
\partial_r A &=& \frac{2}{L}~W(a,b), \label{flow}
\end{eqnarray}
where $W(a, b)$ 
is the same $AdS_4$ superpotential as in (\ref{spoten1}) but now 
is given by a polynomial of $(a,b)$:
\begin{equation}
W(a,b)={1 \over 8}\,a^{3 \over 2}
\dsqrt{(a^2 +7b^2)^2 -112\,(ab-1)}.
\label{spoten}
\end{equation}
The $SO(8)$ gauge coupling constant $g$ in (\ref{RGflow}) has been replaced 
with $\sqrt{2}\,L^{-1}$. 
Note that the derivatives of $W(a, b)$ with respect to $a$ and $b$
do not vanish at the critical points.
However, the $r$-derivatives of $a$ and $b$ do vanish at the critical points.

From the point of view of 11-dimensional supergravity, the flow equations 
(\ref{flow}) can be regarded as an ansatz for the $r$-dependence of 
the geometric parameters $(a,b)$ and the superkink amplitude $A(r)$. 
If the $G_2$-invariant RG flow can be lifted to an $M$-theory solution, 
this ansatz must be correct and the 11-dimensional field equations 
can be solved by calculating all the $r$-derivatives with the flow 
equations (\ref{flow}).

As mentioned in Section 3.2, a certain ansatz for $\th_5$ is required 
to separate out each field strength from the Einstein equation (\ref{einst}). 
Calculating Ricci tensor components by invoking the flow equations 
(\ref{flow}), Eq.\ (\ref{h2h5}) can be written as
\begin{equation}
\left(\frac{12\,a\,\xi^2}{\sin^2 \theta}\right)h_2^2 -c_2\,\th_5^2 
=12\,L^6\,K_0 (a,b)\,a\,\xi^{-2} \sin^6 \theta, \label{h2h5an}
\end{equation}
where the polynomial $K_0$ is given by
$$
K_0 (a,b)=\frac{2\,(ab-1)\,(ab-2)\,(a^2 -7b^2)}{(a^2 +7b^2)^2 -112\,(ab-1)}.
$$
Eq.\ (\ref{h2h5an}) immediately leads to the ansatz:
\begin{eqnarray}
h_2 &=& L^3 \dsqrt{K(a,b)}\,\,\xi^{-2} \sin^4 \theta, \nonumber\\ 
\th_5 &=& 4 L^2 \dsqrt{K_5 (a,b)}\,\,a^{-{1 \over 2}} \sin^2 \theta, 
\label{msol12}
\end{eqnarray}
where the polynomials $K$ and $K_5$ are subject to the constraint
$K-K_5 \equiv K_0$. Note that this ansatz corresponds to (4.13) 
in the previous work \cite{dnw}, except that $\th_5$ is identically zero 
from the beginning in \cite{dnw} which focused on the critical point 
theories only.

The first Maxwell equation (\ref{1stmax}) involves $(W_r, W_\theta)$ 
and seems not to be closed within $(h_2,\th_5)$. 
However, this is not true and by using $H$ given by (\ref{Hh5}) the equation 
can be rewritten as a differential equation closed within $(h_2,\th_5)$. 
Specifically, the equation turns to
\begin{eqnarray}
&& a^3 h_2 D_\theta h_2^{'} +L^2 h_2 D_r \dot{h}_2 
+a^3 \left[2\,f^2 L^2\,a^{-6}\,\xi^{-4}
-\frac{12\,a^{-2}\,\xi^2}{\sin^2 \theta}\right] h_2^2 \nonumber\\
&& \qquad =\frac{1}{2}\left[a^3 (h_2^{'})^2 
+L^2 (\dot{h}_2)^2 -H\right]
+\frac{1}{18}\,c_2^2\left[\frac{1}{L^2}({\cal D}_\theta \th_5)^2 
+\frac{1}{a^3}({\cal D}_r \th_5)^2 \right],
\label{m1stmax}
\end{eqnarray}
where $f$ and $H$ are given by (\ref{f2}) and (\ref{Hh5}) and are directly 
determined by Ricci tensor components without knowing $(W_r, W_\theta)$. 

Applying the ansatz (\ref{msol12}) to the modified equation (\ref{m1stmax}), 
both $K$ and $K_5$ are consistently determined as polynomials of $(a,b)$ 
by using the RG flow equations. They are
\begin{eqnarray}
K(a,b) &=& \frac{1}{4}\,b^2\,(ab-1), \nonumber\\
K_5 (a,b) &=& \frac{1}{4}\,\dfrac{(ab-1)\,(-4a +a^2 b +7b^3)^2}
{(a^2 +7b^2)^2 -112\,(ab-1)}. \label{kk5}
\end{eqnarray}
Substituting $(h_2,\th_5)$ in (\ref{msol12}) with $(K,K_5)$ in (\ref{kk5}) 
into the (44)- and (55)-components of the Einstein equation, one can obtain 
the equations which are closed in $W_r$ and in $W_\theta$, respectively. 
Solving these equations yields
\begin{eqnarray}
W_r &=& -\frac{1}{2L}\,a^2 \Bigl[
a^5 \cos^2 \theta +a^2 b\,(ab-2)\,(4+3\cos 2\theta)
+b^3 \,(7ab-12)\sin^2 \theta \Bigr], \nonumber\\
W_\theta &=& -\frac{a^3 \,
\Bigl[48\,(1-ab)+(a^2 -b^2)\,(a^2 +7b^2)\Bigr]}
{8\,W(a,b)}\,\sin \theta \cos \theta. 
\label{msol34}
\end{eqnarray}
The $(h_2,\th_5)$ in (\ref{msol12}) with $(K,K_5)$ in (\ref{kk5}) 
and the $(W_r,W_\theta)$ in (\ref{msol34}) consist of a $G_2$-invariant 
solution in 11-dimensional supergravity. 
The remaining thing we have to do is to check consistency of this solution 
along the $G_2$-invariant RG flow.

The generalized Freund-Rubin parameter $f$ given by (\ref{f2}) should 
coincide with the $f$ calculated from the solutions $(W_r,W_\theta)$ 
in (\ref{msol34}) through (\ref{freund}). 
One can prove that this is the case 
by using the RG flow equations (\ref{flow}). 
Similarly, the $H$ given by (\ref{Hh5}) is proved to be consistent with 
its definition (\ref{Hdef}) along the RG flow.
All the 11-dimensional field equations (\ref{1stmax}), (\ref{2ndmax}) and 
(\ref{einst}) are satisfied by substituting the 11-dimensional 
solution and imposing the RG flow equations (\ref{flow}). 

We also have to check whether the 11-dimensional solution satisfies the 
boundary conditions (\ref{bc}) at both UV and IR critical points. 
The two input data of $(a,b)$ are
$$
a=1,\quad b=1,
$$
for the $SO(8)$-invariant UV critical point, whereas
$$
a=\sqrt{6 \sqrt{3} \over 5},\quad b=\sqrt{2 \sqrt{3} \over 5},
$$
for the $G_2$-invariant IR critical point. From the $K_5$ in (\ref{kk5}) 
and the $W_\theta$ in (\ref{msol34}), one can easily see that both 
$W_\theta$ and $\th_5$ actually vanish at both critical points. 
Note that at the critical points $W_\theta$ vanishes so that $W_r$ is 
equal to the Freund-Rubin constant $f$. From the $K$ in (\ref{kk5}) 
and the $W_r$ in (\ref{msol34}), two field strengths $(f,h_2)$ read
$$
f =3 L^{-1},\quad h_2 = 0,
$$
at the UV critical point, whereas
$$
f =\frac{108}{25}\sqrt{\frac{2\sqrt{3}}{5}}\,L^{-1},\quad 
h_2 =\frac{\sqrt{2\sqrt{3}}}{10}\,L^3 \,\xi^{-2} \sin^4 \theta,
$$
at the IR critical point to reproduce the critical point theory results 
in \cite{dnw}. 

Thus we have established that 
{\it the solutions (\ref{msol12}), (\ref{kk5}) and (\ref{msol34}) 
actually consist of an exact solution to the 11-dimensional supergravity}, 
provided that the deformation parameters $(a,b)$ of the 
7-ellipsoid and the domain wall amplitude $A(r)$ 
develop in the $AdS_4$ radial 
direction along the $G_2$-invariant RG flow (\ref{flow}). 

Lastly, let us derive the geometric superpotential $\tW$ from the field 
strengths $(W_r,W_\theta)$. Recall that $W_r$ and $W_\theta$ are defined 
by (\ref{fst3}) which is encoded to
\begin{equation}
W_r = \partial_r \tW +3\,(\partial_r A)\,\tW, \qquad
W_\theta = \partial_\theta \tW. \label{s2}
\end{equation}
By integrating $W_\theta$ with respect to $\theta$, the geometric 
superpotential is determined up to a polynomial of vevs such that
$$
\tW = \frac{a^3 \,\Bigl[48\,(1-ab)+(a^2 -b^2)\,(a^2 +7b^2)\Bigr]}
{16\,W(a,b)}\,\cos^2 \theta +X(a,b).
$$
Substituting this equation into the first equation in (\ref{s2}) and 
using the RG flow equations (\ref{flow}), we find that the polynomial 
$X(a, b)$ 
should obey the equation
$$
\partial_r X +\frac{6}{L}\,WX=-\frac{1}{2L}\,a^2 b\,
(-2a^2 +a^3 b -12b^2 +7a b^3).
$$
To solve this equation, we require as an ansatz that the $X(a, b)$ should be 
chosen to make the $\tW(a, b, \theta)$ 
coincide with the 4-dimensional superpotential 
$W(a, b)$ up to a multiplicative constant when 
$\theta$ is fixed to some specific value. 
Then the $X(a, b)$ can be solved as
$$
X(a,b) = \frac{a^3 \,\Bigl[8\,(1 -a b) +b^2 \,(a^2 +7 b^2)\Bigr]}{16\,W(a,b)},
$$
which finally yields the geometric superpotential:
\begin{equation}
\tW(a,b,\theta) =\frac{a^3 \,
\biggl[\Bigl[48\,(1-ab)+(a^2 -b^2)\,(a^2 +7b^2)\Bigr]\cos^2 \theta
+8\,(1 -a b) +b^2 \,(a^2 +7 b^2) \biggr]}{16\,W(a,b)}.
\label{geom}
\end{equation}
Note that the $\tW$ coincides with half the superpotential $W$ 
when $\cos \theta =\sqrt{1/8}$. The similar thing happens in 
the geometric superpotential for the ${\cal N}=2$, $SU(3)\times U(1)$ 
flow found in \cite{cpw,jlp}.

\section{Discussions}

In this paper, we have derived the 11-dimensional Einstein-Maxwell equations 
corresponding to the ${\cal N}=1$, $G_2$-invariant RG flow in the 
4-dimensional gauged supergravity. The 11-dimensional metric generated from 
the nonlinear metric ansatz in \cite{dnw,dn87} is the same as the one 
in the critical point theories except for the non-trivial $AdS_4$ radial 
coordinate($x^4 =r$) dependence of vevs $(\lambda,\alpha)$ which can be 
encoded to the geometric parameters $(a,b)$ for the 7-ellipsoid. 
Provided that the $r$-dependence of the vevs is controlled by the RG flow 
equations, we have found an exact solution to the 11-dimensional field 
equations. With this solution, one can say that the $G_2$-invariant 
holographic RG flow found in \cite{ar2,aw} can be lifted to an ${\cal N}=1$ 
membrane flow in $M$-theory. 
In contrast to the ${\cal N}=2$ membrane flow recently found in 
\cite{cpw,jlp}, our solution does not have any K\"{a}hler structure 
but the almost complex structure on the round ${\bf S}^6$ so that 
the field strength ansatz becomes rather complicated as shown in Section 3.2. 
Moreover, the field strengths $(W_\theta,\th_5)$ must be subject to the 
non-trivial boundary conditions (\ref{bc}) at both UV and IR critical points. 
Nevertheless, the geometric superpotential (\ref{geom}) we have found 
has the same property as in \cite{cpw,jlp}, that is, the geometric 
superpotential becomes half the 4-dimensional superpotential for a specific 
angle of $\theta$ ($\cos \theta = \sqrt{1/8}$ in ${\cal N}=1$ flow). 

For future direction, it may be interesting to study a probe 
membrane moduli space for the ${\cal N}=1$ membrane flow by optimizing 
the geometric superpotential obtained in this paper. It is well known that 
the exceptional group $G_2$ is an automorphism group of octonions \cite{gw}. 
The probe membrane study may shed light on the question how the 
$G_2$ symmetry or octonion algebra plays a role in $d=3$, ${\cal N}=1$ 
superconformal field theory on $M$-theory membranes.

\appendix

\renewcommand{\thesection}{\large \bf \mbox{Appendix~}\Alph{section}}
\renewcommand{\theequation}{\Alph{section}\mbox{.}\arabic{equation}}

\section{\large \bf The 11-dimensional Einstein-Maxwell equations}
\setcounter{equation}{0}

In order to contrast 11-dimensional bosonic equations in this paper 
with the ones in the previous work \cite{dnw}, we will show the 
11-dimensional Einstein-Maxwell equations without specifying the 
$G_2$-covariant tensors on the round ${\bf S}^6$. 
As in Section 3.2 we use Greek indices for the 3-dimensional membrane 
world volume, whereas Latin indices for the $G_2$-invariant 6-dimensional 
unit sphere, say, $\mu,\nu,\dots =1,2,3$ and $m,n,\dots =6,\dots,11$.

The original form of the 11-dimensional Maxwell equation is given 
in (\ref{fieldequations}), that is
$$
\nabla_{\! M} F^{MNPQ} =
-\frac{1}{576}\,\sqrt{-g_{11}}\,\epsilon^{NPQRSTUVWXY}F_{RSTU}F_{VWXY},
$$
where $g_{11}$ stands for the determinant of the 11-dimensional metric 
with the signature $(-+\cdots+)$ and $\nabla_{\! M}$ denotes the covariant 
derivative in 11-dimensions. The epsilon tensor is normalized to be
$\epsilon_{12\dots11}=1$ and $\epsilon^{12\dots11}={g_{11}}^{-1}$. 

Via the field strength ansatz (\ref{fst1}), 
the $(\nu\rho\lambda)$-components of the Maxwell equation become
\begin{equation}
\partial_r (\tilde{e}\,W_r)
+\partial_\theta (g_{44}\,g^{55}\tilde{e}\,W_\theta)
=-\frac{1}{18 L^7}\,\eps6^{\,mnpqrs}
\left[F_{4mnp}F_{5qrs}-\frac{3}{4}F_{45mn}F_{pqrs}\right],
\label{M1}
\end{equation}
where we have defined $\tilde{e}\equiv\Delta^3 \sin^6 \theta$ as before in
Section 3.2. 
The 6-dimensional epsilon tensor is the same as the one in Section 3.2.
The equation (\ref{M1}) goes to the third equation in (\ref{maxf1}) 
and survives as the second Maxwell equation (\ref{2ndmax}). 
Similarly, the $(npq)$-components of the Maxwell equation reduce to
\begin{equation}
\nab6_m F^{mnpq}+\nabla_{\!4} F^{4npq}+\nabla_{\!5} F^{5npq}
=\frac{\Delta}{3 L^7 \sin^6 \theta}\,\eps6^{\,npqrst}
\left(W_\theta F_{4rst}-W_r F_{5rst}\right),
\label{M2}
\end{equation}
where $\nab6_m$ is the covariant derivative on the round ${\bf S}^6$. 
The $T^{npq}$ and the $S^{npq}$ components of this equation turn to 
the first and the second equations in (\ref{maxf1}), respectively, 
and finally reduce to the first Maxwell equation (\ref{1stmax}). 
We also note that (\ref{M2}) corresponds to (3.16) in \cite{dnw}. 
The $(np5)$-components of the Maxwell equation read
\begin{equation}
\nab6_m F^{5mnp}-\nabla_{\!4} F^{45np}
=-\frac{\Delta}{12 L^7 \sin^6 \theta}\,W_r \,\eps6^{npqrst}F_{qrst},
\label{M3}
\end{equation}
which provides the first equation in (\ref{maxf2}). 
Similarly, the $(np4)$-components read
\begin{equation}
\nab6_m F^{4mnp}+\nabla_{\!5} F^{45np}
=\frac{\Delta}{12 L^7 \sin^6 \theta}\,W_\theta \,\eps6^{\,npqrst}F_{qrst}
\label{M4}
\end{equation}
which turns to the second equation in (\ref{maxf2}). 
Other remaining components of the Maxwell equation become 
identically zero and trivially satisfied.

In (\ref{fieldequations}), we find the original form of the 11-dimensional 
Einstein equation:
$$
R_M^{~~N}=-\frac{1}{36}\,\delta_M^{~N}\, F^2 +\frac{1}{3}\,F_{MPQR}F^{NPQR}.
$$
where $F^2 \equiv F_{PQRS}F^{PQRS}$. The 11-dimensional Plank length 
$l_{11}$ has been absorbed into the normalization of gauge field strengths. 
The useful relation is $L=(32\,\pi^2 N)^{1 \over 6}\,l_{11}$ where $L$ 
denotes the radius of the round ${\bf S}^7$ and $N$ is the number of 
coincident $M$-theory membranes.

Applying the field strength ansatz (\ref{fst1}) to the RHS of the 
Einstein equation, $F^2$ becomes
\begin{eqnarray*}
F^2 &=& -24\,\Delta^4 \left( W_r^2 +g_{44}\,g^{55}\,W_\theta^2 \right)
+F_{mnpq}F^{mnpq} \\ &&
+4\,F_{4mnp}F^{4mnp}+4\,F_{5mnp}F^{5mnp}+12\,F_{45mn}F^{45mn},
\end{eqnarray*}
and we obtain
\begin{eqnarray}
R_\mu^{~\nu} &=& \delta_\mu^{~\nu} \biggl[-\frac{1}{36}\,F^2 
-2\,\Delta^4 \left( W_r^2 +g_{44}\,g^{55}\,W_\theta^2 \right) \biggr],
\label{E1}\\
R_4^{~4} &=& -\frac{1}{36}\,F^2 -2\,\Delta^4 \,W_r^2 
+\frac{1}{3}\,F_{4mnp}F^{4mnp}+F_{45mn}F^{45mn}, \label{E2}\\
R_5^{~5} &=& -\frac{1}{36}\,F^2 -2\,\Delta^4 \,g_{44}\,g^{55}\,W_\theta^2 
+\frac{1}{3}\,F_{5mnp}F^{5mnp}+F_{54mn}F^{54mn}, \label{E3}\\
R_m^{~n} &=& -\frac{1}{36}\,F^2 \,\delta_m^{~n}
+\frac{1}{3}\,F_{mpqr}F^{npqr}+F_{4mpq}F^{4npq}+F_{5mpq}F^{5npq}
+2\,F_{mp45}F^{np45},\label{E4}\\
R_5^{~4} &=& -2\,\Delta^4 \,W_r W_\theta + \frac{1}{3}\,F_{5mnp}F^{4mnp}.
\label{E5}
\end{eqnarray}
At the critical points, (\ref{E1}) and (\ref{E2}) combine into (3.5) 
in \cite{dnw}. Similarly, (\ref{E3}) and (\ref{E4}) degenerate into (3.6) 
in \cite{dnw}. (\ref{E5}) has no corresponding equation in \cite{dnw} 
and is characteristic of the domain wall metric. The off-diagonal 
components $R_5^{~4}$ and $R_4^{~5}\equiv g_{44}\,g^{55}R_5^{~4}$ 
also arise in ${\cal N}=2$, $SU(3)\times U(1)$-invariant flow studied 
in \cite{cpw,jlp}.

\section{\large \bf The Ricci tensor in 11-dimensions}
\setcounter{equation}{0}

The $G_2$-invariant 11-dimensional metric (\ref{11dmet}) generated
in Section 3.1 can be rewritten as
\begin{equation}
ds_{11}^2 =a\,\xi^{4 \over 3}\left(dr^2 +e^{2 A(r)}
\eta_{\mu\nu}dx^\mu dx^\nu \right)+
L^2 \,\xi^{-{2 \over 3}}\left(a^{-2}\,\xi^2 \,d\theta^2 
+\sin^2 \theta\,d\Omega_6^2\right),
\label{11dmet2}
\end{equation}
with the quadratic form $\xi^2=a^2 \cos^2 \theta +b^2 \sin^2 \theta$ 
in (\ref{xi}). 
Imposing $r$-dependence to two vevs $(a,b)$, the warped metric 
(\ref{11dmet2}) generates Ricci tensor components:
\begin{eqnarray}
R_1^{~1} &=& \frac{1}{6L^2}\,a^{-3}\xi^{-{10 \over 3}}
\biggl[3L^2 \xi^2 \left[
\dot{a}^2 -3a\dot{a}\dot{A}
-a\ddot{a}-2a^2 \left(\ddot{A}+3\dot{A}^2 \right)
\right] \nonumber\\ &&
+4a^2 \left(a^3 \xi^{'\,2} +L^2 \dot{\xi}^2 \right)
-4a^2 \xi \left[ 
a^3 \left(\xi^{''}\!+6\,\xi^{'}\!\cot\theta\right) 
+L^2 \left(\ddot{\xi}+3\,\dot{\xi}\dot{A}\right) 
\right] \biggr],\\ 
R_4^{~4} &=& \frac{1}{6L^2}\,a^{-3}\xi^{-{10 \over 3}}
\biggl[-3L^2 \xi^2 \left[
2\dot{a}^2 +3a\dot{a}\dot{A}
+a\ddot{a}+6a^2 \left(\dot{A}^2 +\ddot{A}\right)
\right] \nonumber\\ &&
+4a^2 \left(a^3 \xi^{'\,2} -2L^2 \dot{\xi}^2 \right)
-4a^2 \xi \left[ 
a^3 \left(\xi^{''}\!+6\,\xi^{'}\!\cot\theta\right) 
+L^2 \left(\ddot{\xi}+3\,\dot{\xi}\dot{A}\right) 
\right] \biggr],\\
R_5^{~5} &=& \frac{1}{3L^2}\, a^{-3}\xi^{-{10 \over 3}}
\biggl[-3L^2 \xi^2 \,\dot{a}^2 
+3L^2 a\,\xi^2 \left(\ddot{a}+3\dot{a}\dot{A}\right)
\nonumber\\ &&
+a^5 \left[
18 \xi^2 -4\xi^{'\,2}
-2\xi \left(\xi^{''}\!-12\,\xi^{'}\!\cot\theta\right)
\right]
-2L^2 a^2 \left[
-\dot{\xi}^2 +\xi\left(\ddot{\xi}+3\,\dot{\xi}\dot{A}\right)
\right] \biggr],\\
R_6^{~6} &=& \frac{1}{3L^2}\,a^{-1}\xi^{-{10 \over 3}}
\Biggl[\frac{15 a\,\xi^4}{\sin^2 \theta}
-a^3 \left[
\frac{3\left(2+3\cos 2\theta\right)\xi^2}{\sin^2 \theta}
+\xi^{'\,2} -\xi\left(\xi^{''}\!+6\,\xi^{'}\!\cot\theta\right)
\right] 
\nonumber\\ &&
+L^2 \left[-\dot{\xi}^2 
+\xi\left(\ddot{\xi}+3\,\dot{\xi}\dot{A}\right)\right]
\Biggr],\\
R_5^{~4} &=& a^{-2}\xi^{-{10 \over 3}}
\biggl[6\left(a\,\xi\,\dot{\xi}-\dot{a}\,\xi^2\right)\cot\theta
-2a\,\dot{\xi}\,\xi^{'}\biggr],
\end{eqnarray}
with identities:
$$
R_2^{~2}=R_3^{~3}=R_1^{~1},\quad 
R_7^{~7}=R_8^{~8}=R_{9}^{~9}=R_{10}^{~10}=R_{11}^{~11}=R_6^{~6},\quad 
R_4^{~5}=L^{-2}\,a^{3} \,R_5^{~4}.
$$
We have used the abbreviation such as 
$\dot{a}\equiv\partial_r a$, $\dot{\xi}\equiv\partial_r \xi$, 
$\xi^{'}\equiv\partial_\theta\,\xi$ and so on.
By using for example the relations
$$
\xi\,\xi^{'} = (b^2 -a^2) \cos\theta \sin\theta,\qquad
\xi\,\dot{\xi} = a\dot{a}\,\cos^2 \theta+b\dot{b}\,\sin^2 \theta,
$$
the Ricci tensor components are given by $(a,b,A)$ 
and their derivatives only.

Applying the RG flow equations (\ref{flow}), all the $r$-derivatives in the 
Ricci tensor components can be replaced with polynomials of $(a,b)$.
As a simple calculation, let us derive the polynomial $K_0$ in Section 3.3.
The linear combination in the RHS of (\ref{h2h5}) can be evaluated as
$$
R_1^{~1}+2 R_6^{~6} = \frac{1}{2 L^2}\,a^{-3}\xi^{-{4 \over 3}}
\biggl[4 a^5 +20 a^3 b^2 +L^2 \dot{a}^2 
-L^2 a \left(\ddot{a}+3\dot{a}\dot{A}\right)
-2 L^2 a^2 \left(\ddot{A}+3\dot{A}^2 \right)\biggr]. 
$$
Iterative use of the RG flow equations in the RHS yields
$$
R_1^{~1}+2 R_6^{~6} = \dfrac{64\,(ab-1)\,(ab-2)\,(a^2 -7b^2)}
{(a^2 +7b^2)^2-112\,(ab-1)}\,L^{-2}\xi^{-{4 \over 3}}.
$$
from which one can read the polynomial $K_0$ in (\ref{h2h5an}) 
as in the text.

\vspace{1cm}
\centerline{\bf Acknowledgments}

This research was supported by 
grant 2000-1-11200-001-3 from the Basic Research Program of the Korea 
Science $\&$ Engineering Foundation. 
We are grateful to H. Nicolai for correspondence.



\end{document}